\documentclass[manuscript]{aastex}

\usepackage{pdfsync}

\shorttitle{KHI in the Solar Corona}
\shortauthors{M\"ostl et al.}

\begin{document}

\title{The Kelvin-Helmholtz Instability at CME-Boundaries in the Solar Corona: Observations and 2.5D MHD Simulations}

\author{U.V. M\"ostl, M. Temmer, and A.M. Veronig}
\affil{IGAM-Kanzelh\"ohe Observatory, Institute of Physics, University of Graz, 8010 Graz, Austria}
\email{ute.moestl@uni-graz.at}


\begin{abstract}
The Atmospheric Imaging Assembly onboard the Solar Dynamics Observatory observed a coronal mass ejection with an embedded filament on 2011 February 24, reavealing quasi-periodic vortex-like structures at the northern side of the filament boundary with a wavelength of approximately $14.4$ Mm and a propagation speed of about $310 \pm 20$ km s$^{-1}$. These structures could result from the Kelvin-Helmholtz instability occurring on the boundary. We perform $2.5$D numerical simulations of the Kelvin-Helmholtz instability and compare the simulated characteristic properties of the instability with the observations, where we obtain qualitative as well as quantitative accordance. We study the absence of Kelvin-Helmholtz vortex-like structures on the southern side of the filament boundary and find that a magnetic field component parallel to the boundary with a strength of about $20 \%$ of the total magnetic field has stabilizing effects resulting in an asymmetric development of the instability. 
\end{abstract}



\section{Introduction}

For the dynamics of the solar atmosphere, magnetohydrodynamic (MHD) instabilities are thought to play a major role \citep[e.g.][]{ryutova10,foullon11,ofman11,taroyan11,innes12}. Specifically, there are boundary layers, where the Kelvin-Helmholtz instability (KHI) might be able to develop, as treated in several theoretical studies on this topic \citep[e.g.][]{heyvaerts83,rae83,karpen93,andries01,kolesnikov04,soler10,zaqarashvili10,soler12}. In principal, the KHI arises at boundaries with a velocity shear parallel to a boundary layer \citep{chandrasekhar61}, and there exist many plasma configurations in space where the KHI can occur \citep[e.g.][and references therein]{miura97}.

With the unprecedented observations by the Atmospheric Imaging Assembly \citep[AIA;][]{lemen12} onboard Solar Dynamics Observatory \citep[SDO;][]{pesnell12} it is possible to gain observational evidence for instabilities in the solar atmosphere. SDO/AIA carries four telescopes providing narrow-band full-disk imaging of the Sun in different UV and EUV passbands at a cadence as high as $12$ s and spatial resolution of $1''.5$ \citep{lemen12}. \cite{foullon11} and \cite{ofman11} were the first to present observations of the KHI associated with a coronal mass ejection (CME). \cite{ofman11} analyzed a series of vortices seen along the boundary of an evolving coronal dimming region on 2010 April 8. They compared for the first time the observational features with the results of a 2.5D MHD model of the KHI and found good qualitative agreement. \cite{foullon11} reported observations of wavy vortex-like structures on one flank of a CME that occurred on 2010 November 3. These studies hint at the existence of the KHI in the solar corona. Yet, we need to have more observations and theoretical (numerical) studies on this subject to be clearly able to distinguish between structures produced by the KHI and such that have a different origin.

In this letter, we report quasi-periodic vortex-like structures observed on the boundary of an erupting filament/CME on 2011 February 24 in SDO/AIA high cadence imagery. We follow a similar approach as \cite{ofman11} and give a detailed comparison of these observations and a 2.5D MHD simulation of the KHI. Qualitative as well as quantitative analyses of both spatial and temporal features are presented. Furthermore, we investigate the effect of a magnetic field component parallel to the boundary layer on the evolution of the instability and determine a threshold for the field strength needed to stabilize the boundary. 

\section{Observations and Analyses}\label{sec1}

\begin{figure}
\epsscale{1.0}
\plotone{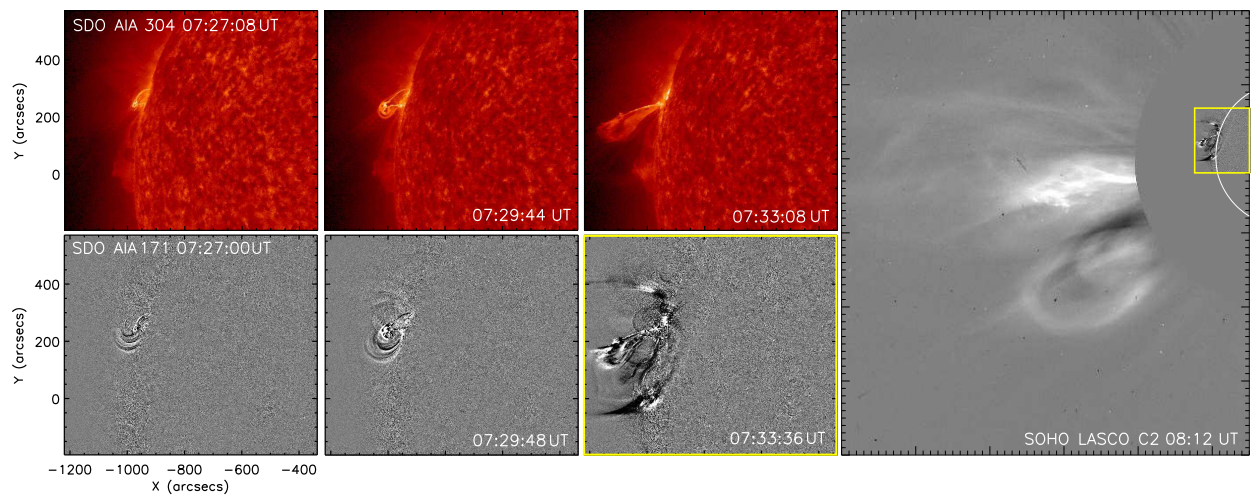}
\caption{Left: Observations of SDO/AIA of the erupting filament. The $304\,\mbox{\AA}$ channel shows direct images (movie in online version), the $171\,\mbox{\AA}$ channel the corresponding running difference images. Right: Observations of SOHO/LASCO of the associated CME. The yellow rectangle shows the AIA $171\,\mbox{\AA}$ running difference image about $30$ minutes prior to the LASCO observations. \label{fig:SDOobsKine}}
\end{figure}

On 2011 February 24, a CME eruption with an embedded filament was observed by the AIA instrument onboard SDO in association with a GOES M3.5 limb flare \citep{battaglia11,martinez12}. The top three panels of Figure \ref{fig:SDOobsKine} show direct images in the AIA $304\,\mbox{\AA}$ passband (dominated by emission from He {\small II}; $\log{(T)} = 4.7$), the bottom panels display running difference images in the $171\,\mbox{\AA}$ channel \citep[dominated by emission from Fe {\small IX}; $\log{(T)} = 5.8$;][]{lemen12}. On the right panel we show the associated CME observed by the Large Angle and Spectrometric Coronagraph (LASCO) C2 onboard the Solar and Heliospheric Observatory (SOHO). The insert (yellow rectangle) represents the AIA $171\,\mbox{\AA}$ running difference image recorded about $30$ minutes before the LASCO image shown. The CME structure breaks up rather early in its evolution (around 07:30:00 UT), and quasi-periodic vortex-like structures develop around 07:36:00 UT on the northern side of the boundary of the filament embedded in the CME (see accompanying movie in the online version of this publication). 

\begin{figure}
\epsscale{1.0}
\plotone{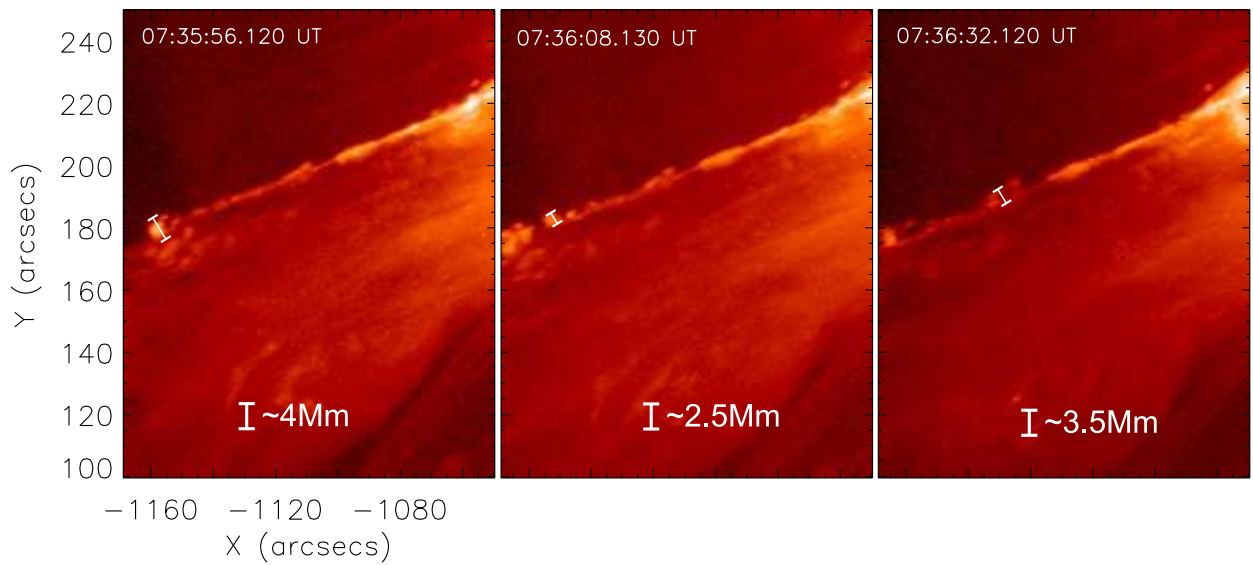}
\caption{Zoom into the AIA $304\,\mbox{\AA}$ images, showing the northern boundary of the filament and indicating the height of the observed structures. \label{fig:Height}}
\end{figure}

The vortex structures are best observed in the AIA $304\,\mbox{\AA}$ passband ($T \approx 0.5$ MK, see Figures \ref{fig:SDOobsKine} and \ref{fig:Height}). Wavebands imaging the outer corona in temperatures between $0.4$ MK - $2$ MK ($131\,\mbox{\AA}$, $171\,\mbox{\AA}$, $193\,\mbox{\AA}$ and $211\,\mbox{\AA}$) show only the parts of the vortices close to the boundary layer.  

\begin{figure}
\epsscale{.9}
\plotone{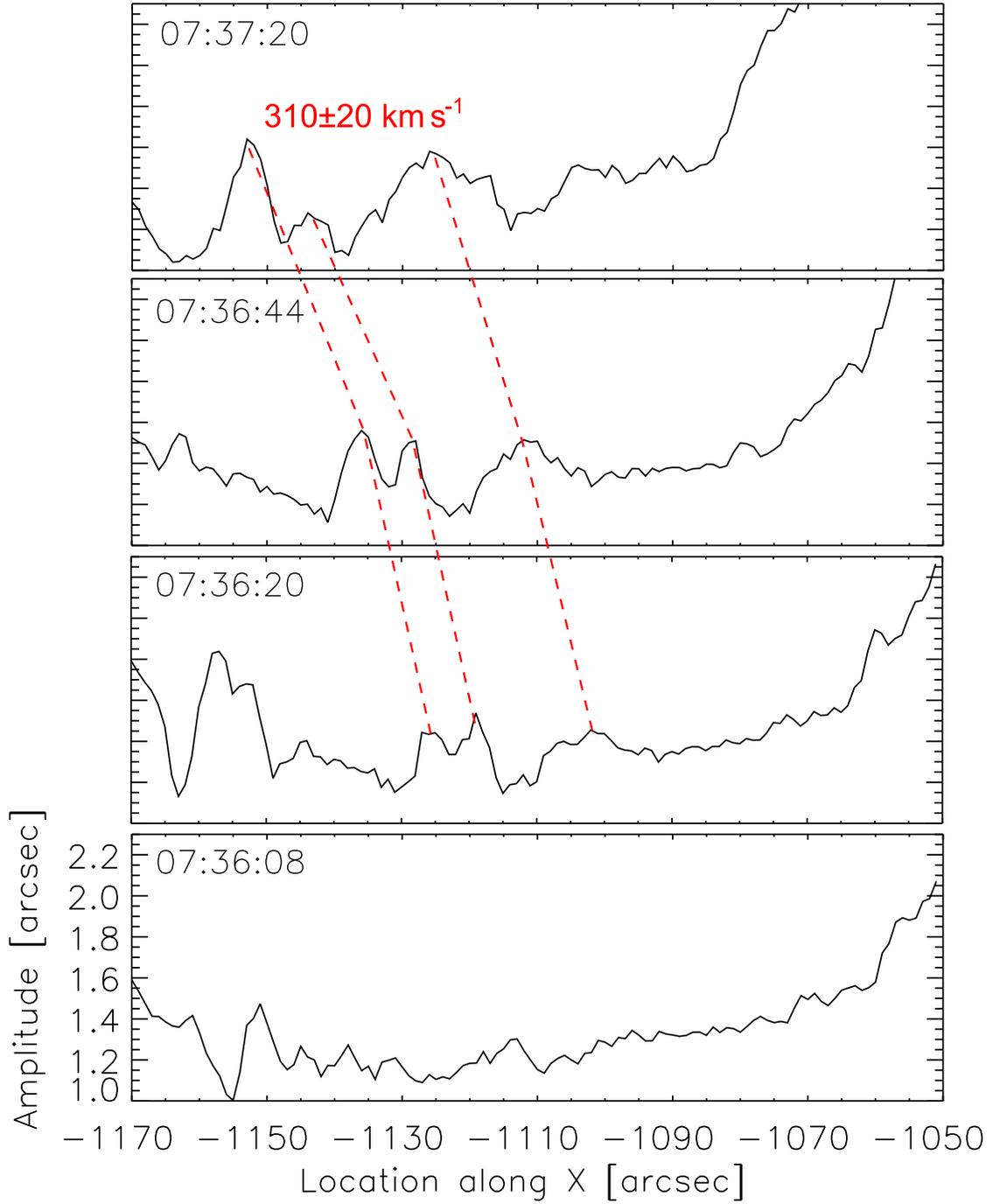}
\caption{Profiles of the quasi-periodic structures along the northern boundary layer of the filament for different times, measured from direct images in the AIA $304\,\mathring{A}$ channel. The time is given in UT. The propagation speed of these structures is about $310 \pm 20$ km s$^{-1}$.  \label{fig:Period}}
\end{figure}

Measuring the front of the filament, we derive an average speed of $\approx 470$ km s$^{-1}$. The height of the structures lies in the range of $2.5$ to $4$ Mm (Figure \ref{fig:Height}). In Figure \ref{fig:Period} we show cuts along the northern edge of the filament for different times. These plots clearly reveal the quasi-periodic appearence of the vortex-like structures, their evolution and propagation with time, which allows us to estimate characteristic quantities from these observations. The spatial period of the vortex-like structures is found to be approximately $20''$ (Figure \ref{fig:Period}), which corresponds to a wavelength $\lambda$ of approximately $14.4$ Mm. The average propagation speed $v_{\rm{prop}}$ of the observed structures along the boundary is about $310\,\pm20$ km s$^{-1}$. Since this event occurred on the limb of the Sun, we can neglect projection effects.

\section{Simulation Setup}

We solve the ideal MHD equations numerically with the so-called TVD Lax-Friedrichs scheme \citep{toth96}. The numerical algorithm was already used in previous studies on the Kelvin-Helmholtz instability \citep{amerstorfer10,moestl11,zellinger12}.

We estimate the plasma parameters in the corona using typical values, which can be found in the literature \citep{aschwanden04}. In the corona, the electron density $n_e$ is taken to be $1 \times 10^{15}$ m$^{-3}$, the plasma pressure $p$ is $0.09$ Pa, and the magnetic field $B$ is $10 \times 10^{-4}$ T. With these values, we obtain for the total pressure $\Pi = 0.5$ Pa, which is the sum of plasma and magnetic pressures. In the filament, we assume that the density is at least $10$ times higher than the corona value and that $T \leq 1 \times 10^5$ K. With further assumptions that $n_i \approx n_e = n$ and $T_i \approx T_e = T$ we have for the plasma pressure $p = 2\,n\,k\,T = 0.028$ Pa. The subscript $i$ denotes the values for the ions. We take $450$ km s$^{-1}$ as the approximate speed of the filament $v_f$ and assume that the surrounding coronal plasma moves at half of this speed. These plasma parameters are used as input for the simulations. 

The general setup for the simulations is such that the $x$-direction is along the boundary layer and the $y$-direction is perpendicular to it. This means that the plasma parameters change in $y$-direction, and the disturbances (vortices) propagate along the $x$-direction. The background plasma parameters vary according to a $\tanh$-profile. Thus, we do not have a tangential discontinuity, but the boundary layer has some finite thickness. 

As initial condition we need an equilibrium plasma configuration. In ideal MHD simulations, this means that the total pressure $\Pi$ has to be constant across the boundary layer, i.e.~in our case
\begin{equation}
\frac{\partial \Pi}{\partial y} = 0\,,
\end{equation}
with
\begin{equation}
\Pi = p + \frac{B^2}{2\,\mu_0}\,,
\end{equation}
where $p$ is the thermal pressure, $B$ is the magnetic field and $\mu_0$ is the permeability in vacuum. In order to have an initial equilibrium, we determine the total and the plasma pressure from the plasma background parameters and then calculate the magnetic field according to
\begin{equation} \label{equ:magField}
B = \sqrt{2\,\mu_0\,(\Pi - p)}\,.
\end{equation}
For the cases, where we have two magnetic field components, $B_x$ and $B_z$, we assume a constant $B_x$ and then use Equation \ref{equ:magField} to calculate $B_z$. Onto this equilibrium we impose small random initial disturbances of $v_x$ and $v_y$, with an amplitude of $1\%$ of the initial filament velocity.

We use normalized quantities in the numerical simulations. The normalization constants are the initial filament speed $v_f$, the filament mass density $\rho_f$ and the half-width of the boundary layer $a$. Accordingly, the time is then normalized with $a/v_f$, the pressures with $\rho_f\,v_f^2$ and the magnetic field with $\sqrt{\mu_0\,\rho_f\,v_f^2}$. The computational box extends from $0$ to $50\,a$ in $x$-direction and from $-25\,a$ to $25\,a$ in $y$-direction. At the $x$-boundaries we impose periodic boundary conditions and at the $y$-boundaries transmissive ones.

\section{Results} 

\begin{figure}
\epsscale{1.0}
 \plotone{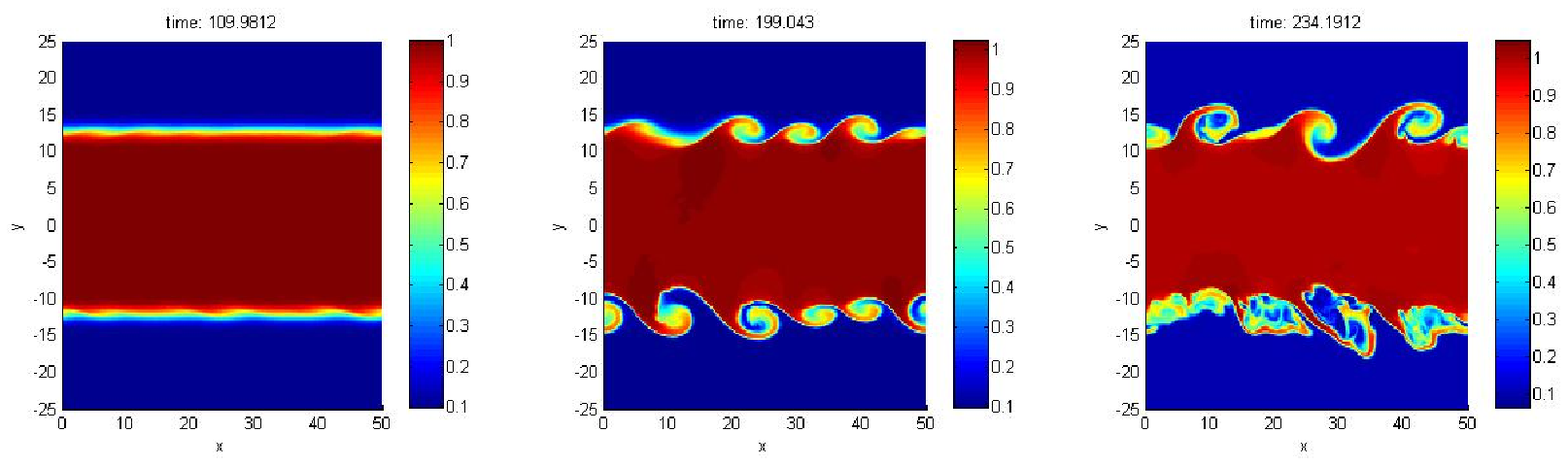}
 \plotone{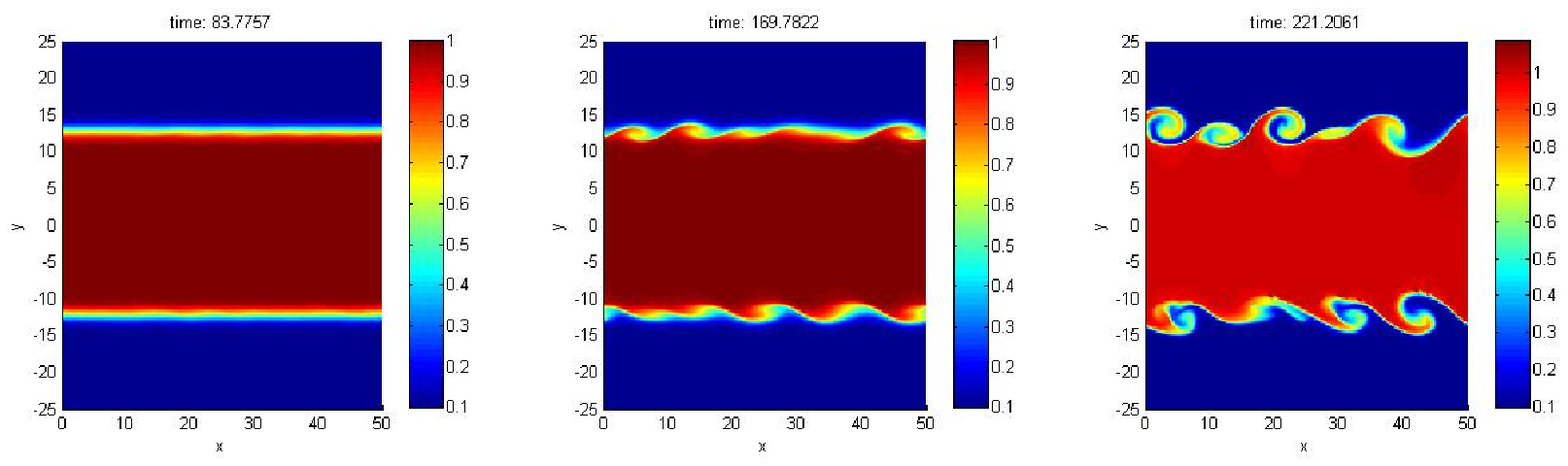}
 \plotone{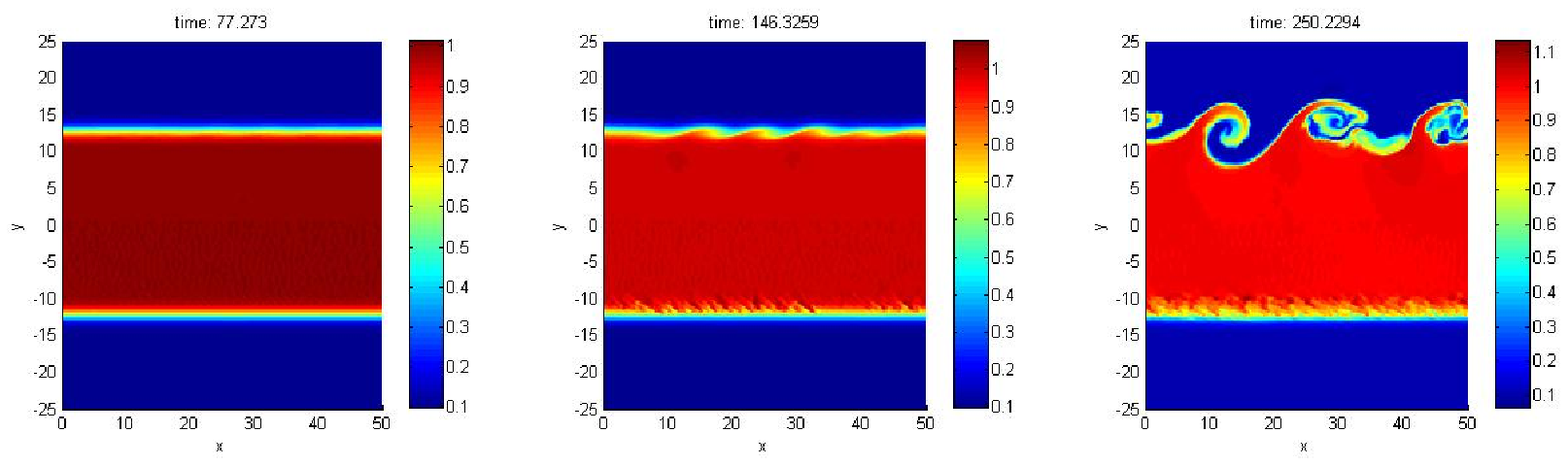}
\caption{Time series of the mass density (color code). The plasma flows from right to left. Top panel: Only a $z$-component of the magnetic field was assumed. Middle panel: An $x$-component of the magnetic field of $10$\% of the total magnetic field strength was assumed. Bottom panel: An $x$-component of the magnetic field of $20$\% of the total magnetic field strength was assumed (movies in online version).}
\label{fig:densSeries_Bz_Bx}
\end{figure}

The top panel of Figure \ref{fig:densSeries_Bz_Bx} shows the (normalized) mass density for different (normalized) times in the simulation run. The blue layer represents the surrounding coronal plasma, whereas the red layer is the ejected filament. For this simulation run, only a $z$-component of the magnetic field was taken on both boundaries. The $x$- and $y$-components of $B$ are zero, as well as the $z$-component of the velocity. The plasma flows from right to left. After some time, small initial perturbations grow in amplitude and eventually reach their nonlinear vortex phase. Then, a saturation sets in and the regular vortex-structures become irregular, as does the whole boundary layer. At $t \approx 200$ the vortices have fully formed on both boundary layers. However, the observations show vortices only on one side of the filament boundary. 

From theory it is known that a magnetic field parallel to the boundary layer can have a stabilizing effect on the KHI due to magnetic field line tension. The Alfv\'en Mach number $M_A$, determined only with the parallel magnetic field component, has to be larger than $2$ for the instability to develop. Combining the effects of compressibility and magnetic field line tension results in the instability condition that $2\,v_A < \Delta v \leq 2\,v_s$, where $\Delta v$ is the total velocity jump across the boundary layer, $v_A$ is the Alfv\'en velocity and $v_s$ the sound velocity, or equivalently $v_A < v_s$ \citep[for the exact derivation of the instability criterion we refer to][]{miurapritchett82}. We performed simulations including an $x$-component of the magnetic field at the lower boundary layer to derive the field strength needed to stabilize the boundary layer.

The middle and bottom panels of Figure \ref{fig:densSeries_Bz_Bx} show the results of two simulation runs where an $x$-component of the magnetic field is included at the lower boundary layer. The middle panel is the result with $B_x = 0.1\,B$ and the bottom panel with $B_x = 0.2\,B$. For the first case, there are still vortices visible at the lower boundary, whereas for the latter case, the parallel magnetic field is large enough to stabilize the boundary layer. This stabilizing effect of a strong parallel $B$ \citep{chandrasekhar61} is a possible reason why the vortices are only observed on one side of the erupting filament.

From the simulation that most closely resembles the observations (bottom panel of Figure \ref{fig:densSeries_Bz_Bx}), we determine the wavelength $\lambda$ (i.e.~the distance between two vortices) and the height $h$ of the vortex-structure. We derive an approximate $\lambda$ of about $10$ to $15\,a$ and an approximate $h$ of about $3$ to $5\,a$.

In order to get physical values, we have to make an estimation of the half width of the shear layer $a$. From the observations shown in Figure \ref{fig:Height}, we can estimate the boundary layer to be about $1$ to $2$ Mm, resulting in $a \approx 0.5 - 1$ Mm. Taking these values, the wavelength and the height of the simulated structures yield (i) $\lambda \approx 5 - 7.5$ Mm and $h \approx 1.5 - 2.5$ Mm, or (ii) $\lambda \approx 10 - 15$ Mm and $h \approx 3 - 5$ Mm, respectively. The propagation velocity $v_{\rm{prop}}$ of the simulated KH vortices lies around $330$ km s$^{-1}$. All of these characteristic features of the simulated instability are in basic quantitative agreement with the observations ($\lambda \approx 14.4$ Mm, $h \approx 2.5 - 4$ Mm, $v_{\rm{prop}} \approx 310 \pm 20$ km s$^{-1}$). Furthermore, we find that a $B_x$ of approximately $2 \times 10^{-4}$ T is needed to stabilize the lower boundary layer.

\begin{figure}
\epsscale{0.9}
 \plotone{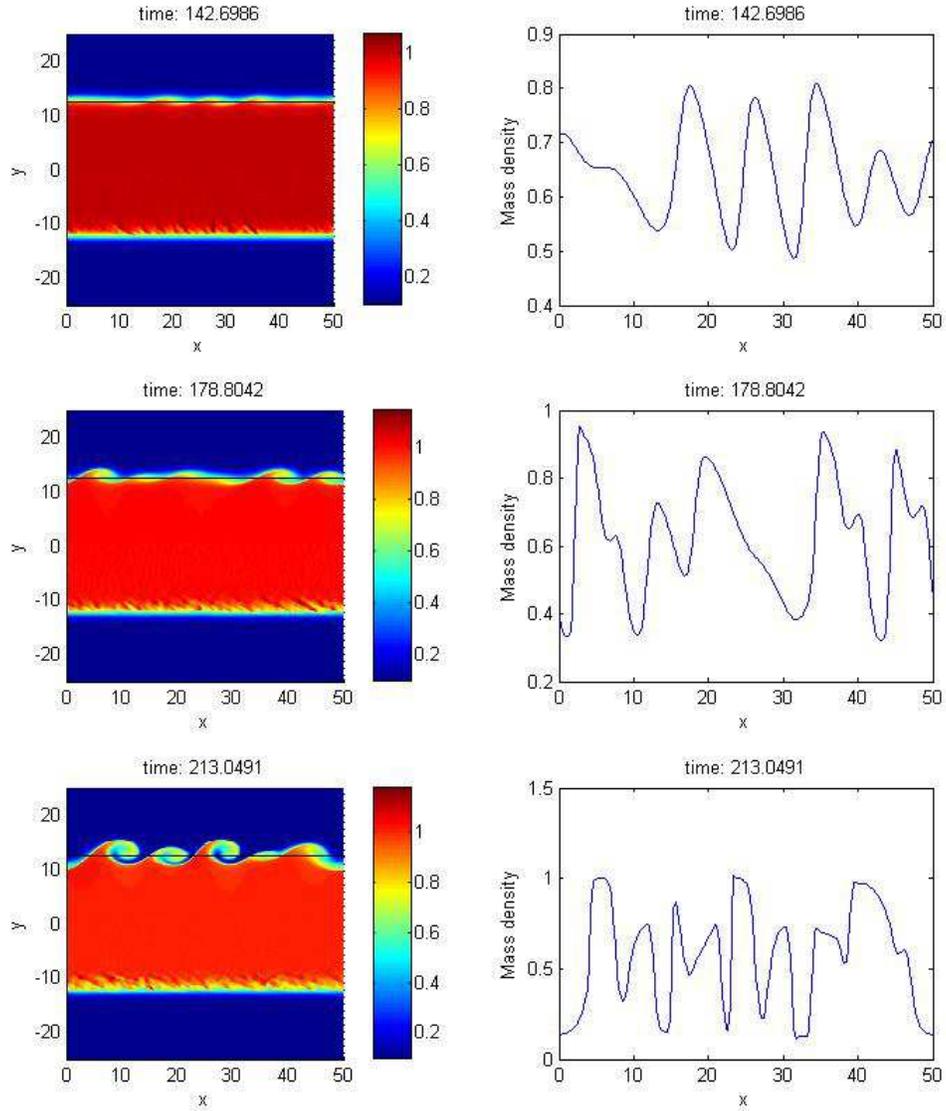}
 \caption{Left: Evolution of the KH vortices. The black lines indicate, where the cuts shown on the right plots were taken. Right: Profiles of the density through the vortices at different time steps.}
\label{fig:VortProf}
\end{figure}

Qualitatively, we can compare the structure of the simulated vortices with the observations. Figure \ref{fig:VortProf} shows on the left side the evolution of the simulated KH vortices and on the right side the profiles through the vortices for each time step. A rather characteristic double peak appears when the vortices are rolled-up. Such double peaks are also visible in the profiles obtained from the direct images presented in Figure \ref{fig:Period}. Another feature associated with an instability is a growing amplitude of the perturbations, which is seen in both, simulations and observations (Figures \ref{fig:VortProf} and \ref{fig:Period}, respectively).

\section{Discussion and Conclusions}

There are various important implications resulting from the occurrence of the KHI at the boundaries of CMEs. As \citet{foullon11} discuss, the KHI at CME surfaces may affect the drag force \citep{vrsnak12}, and hence, the CME kinematics in interplanetary space and its geoeffectiveness. Due to reconnection inside KH vortices, a plasma transport across the boundary can be initiated \citep[e.g.][]{nykyri04} leading to a mass loss of the CME. Additionally, magnetic reconnection can lead to the erosion of the initial magnetic flux, again influencing the geoeffectiveness of CMEs \citep{dasso07,taubenschuss10,ruffenach12}.

We present numerical simulations of the KHI on a boundary of a filament embedded in a CME. We find qualitative as well as quantitative agreements between the simulation results and the observed features of quasi-periodic vortex-like structures observed at the northern boundary of a filament during an eruption on 2011 February 24. Additionally, we have shown that a magnetic field component parallel to the boundary layer with a strength of about $0.2\,B$, corresponding in our case to $2 \times 10^{-4}$ T, results in a stabilization of the boundary layer and could be an explanation of the observed asymmetric development of the vortex-like structures. Our assumption of a tenfold density jump towards the filament is a very conservative one, as one can expect a larger density increase of about $100$ or more. As was shown in previous studies \citep{amerstorfer10,moestl11}, a larger density ratio results in smaller instability growth rates and, thus, it exhibits a stabilization of the boundary layer. Contrary to this stabilizing effect, a larger velocity shear than the one assumed in this case study would act destabilizing. 

The numerical MHD model applied is of course an idealized one with simplifying assumptions of the geometry of the CME and of the parameters. The model uses standard input plasma parameters, which are typical for the observed features but may differ for this specific event. Nevertheless, it captures important basic physical properties of the studied boundary layer processes. We point out that we are limited in observations coming from a single vantage point. The vortex structures might be affected by projection effects and possibly signatures of twisted field lines (flux rope), viewed from aside, may show similar structures.

The stabilizing effects might be large enough, such that the KHI is only seldomly able to develop at CME boundaries in the solar corona. For the present case study we found evidence for the occurrence of the KHI on boundary layers in the solar corona. More observational data of possible KHI formation are needed to better understand its physics and its consequences on propagating MHD structures. 

\acknowledgments{UVM, MT and AMV gratefully acknowledge the Austrian Science Fund (FWF): P21051-N16, V195-N16 and P24092-N16.}

{\it Facilities:} \facility{SDO (AIA)}

\end{document}